\documentclass[11pt]{article}

\usepackage{moriond}
\usepackage[dvips]{graphicx}
\usepackage{subfigure}

\def\be{\begin{equation}}
\def\ee{\end{equation}}
\def\bea{\begin{eqnarray}}
\def\eea{\end{eqnarray}}

\newcommand{\bk}{{\bf k}}

\begin{document}
\vspace*{4cm}
\title{CENTRAL EXCLUSIVE DIJET PRODUCTION}

\author{A. DECHAMBRE$^1$, J.R. CUDELL$^1$, O.F. HERN\'ANDEZ$^{2}$, I. P. IVANOV$^1$}

\address{~\\ $^1$IFPA, AGO Dept., University of Li\`ege, Li\`ege, 
Belgium\\
$^2$Physics Dept., McGill University and Marianopolis College, Montr\'eal, Qu\'ebec, Canada\\}
\maketitle\abstracts{Calculations of central exclusive production
are affected by very large perturbative and non-perturbative corrections.
In this talk, we summarize the results of a study of the uncertainties on these corrections in the case of exclusive dijet production.}
\section{Central Exclusive Calculation and Data}
\begin{figure}
\centering{\includegraphics[width=8.4cm,keepaspectratio]{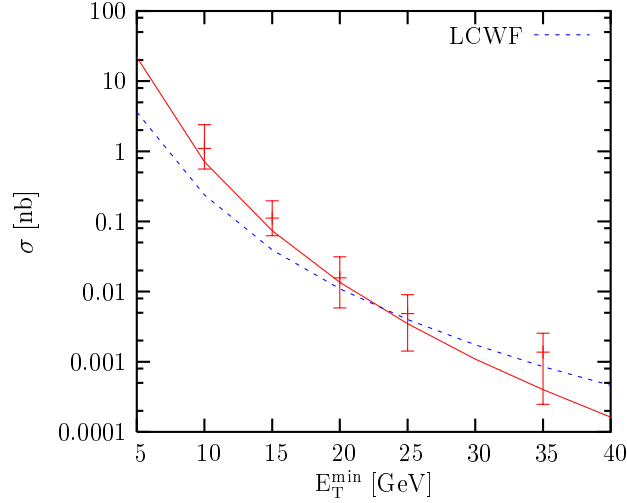}}
\caption{\label{impact}The CDF Run II central exclusive dijet cross section as a function of the minimum transverse energy of the jets and a reference curve corresponding to parameters chosen so that it goes through the CDF Run II data. Sensitivity of the cross section to the impact factor and data. The full line corresponds to an unintegrated gluon density and LCWF corresponds to an impact factor where the partons wave functions are parametrized on the light-cone.}
\end{figure}

At the end of 2007, the CDF collaboration working at the Tevatron $p\bar{p}$ collider published~\cite{Aaltonen:2007hs} data on exclusive dijet events. This kind of events is characterized by a large rapidity gap between the two-jet system and the remaining proton and antiproton. Such hadronic diffractive processes, 
\begin{equation}
p+\bar{p} \to p + gap + X + gap + \bar{p},
\end{equation}
are dominated at high energy by the exchange of a colour singlet trajectory with the quantum numbers of the vacuum, the pomeron. Practically, and contrarily to inelastic processes where each particle is hidden in numerous products of the reaction, the final proton and antiproton (or the remnants) are separated in the forward detectors and the system centrally produced is isolated. The CDF central exclusive dijet cross section is plotted in Fig.~\ref{impact}
as a function of the minimum transverse energy of the jets. It is a rather small cross section but it corresponds to 2.10$^5$ exclusive dijet events at the Tevatron Run~II luminosity. This mode of production is interesting because, from the measurement of the momenta of the daughter proton and antiproton, it is possible to reconstruct the mass of the centrally-produced system without detecting it. It could be a useful process to study a Standard Model Higgs boson with a mass close to 120~GeV at the LHC.

\begin{figure}
\centering{\includegraphics[width=14.2cm,keepaspectratio]{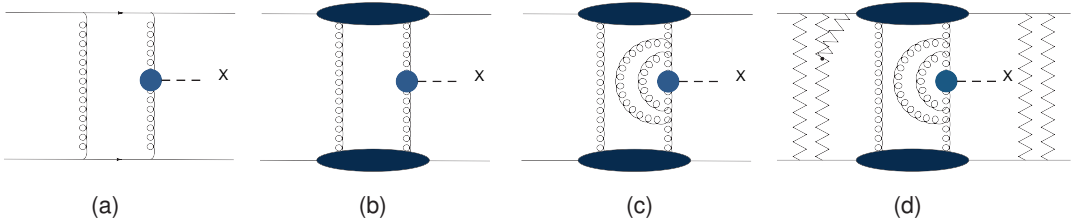}}
\caption{\label{pieces} Steps of a central exclusive production calculation: (a) Lowest order QCD calculation, (b) Impact factor, (c) Virtual corrections, (d) Screening corrections.}
\end{figure}

\begin{figure}
\centering{\includegraphics[width=5cm,keepaspectratio]{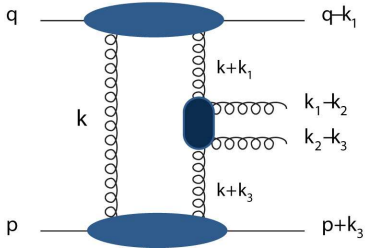}}
\caption{\label{diagram} General diagram and kinematics.}
\end{figure}

Typical calculations~\cite{Khoze:2000cy,Bzdak:2004ux,Berera:1995vi} of central exclusive dijet production can be divided into four pieces. They begin with the lowest-order QCD calculation, in which two partons exchange two gluons and the final state is a colour singlet as in Fig.~\ref{pieces}.a. The amplitude can be analyticaly calculated but is not infrared finite before embedding the partons into the proton via an impact factor~\footnote{The impact factor models the behavior of a real proton and goes to zero when the gluons become very soft~\cite{Gunion:1976iy}. It is sometimes called the proton form factor.}, as in Fig.~\ref{pieces}.b. Next, we have to add the virtual corrections that are large and viewed as a Sudakov form factor, see Fig.~\ref{pieces}.c. And finally, we must take into account proton and antiproton rescatterings as in Fig.~\ref{pieces}.d. This final piece is called the gap survival probability.
While the first part of the calculation is under theoretical control, the second part, including the impact factor, the Sudakov form factor and the gap survival, is not and can lead to large uncertainties. In the following, we shall detail each piece and show the sensitivity of the exclusive dijet cross section to it~\cite{CDHI}. Our notations are given in the diagram shown in Fig.~\ref{diagram} where $\bk_{1}$ and $\bk_{3}$ are the momenta transferred to the proton whereas the relative momentum between the two produced gluons is $\sim 2\bk_{2}$ and is related to the mass of the gluon system.
\section{Uncertainties}
The amplitude is dominated by its imaginary part and can be calculated analytically. Its square is given by 
\begin{equation}
(\mathrm{Im} \mathcal{M})^2\propto \frac{g^{12}}{(\bk^2_2)^2}\int\frac{d^2\bk d^2\bk'}
{\bk^2(\bk+\bk_{1})^2(\bk+\bk_{3})^2\bk'^2(\bk'+\bk_{1})^2(\bk'+\bk_{3})^2}\times 
(C_{0}|M_{0}|^2+C_{2}|M_{2}|^2).
\end{equation}
It is of order $\alpha_s^6$ and falls quickly with $\bk_{2}$, i.e.~with 
the mass of the gluon system. The expression includes the gluon 
propagators and a piece 
coming from the sub-amplitude $gg$~$\to$~$gg$.
\begin{equation}
|M_{0}|^2=1,\quad |M_{2}|^2=\frac{u_{gg}^4+t_{gg}^4}{s_{gg}^4},
\end{equation}
are the sub-amplitudes where the total helicity of the two-gluon sub-process is respectively equal to zero or two, while $C_{0}$ and $C_{2}$ 
are certain products of $\bk$, $\bk'$, $\bk_1$ and $\bk_3$. This is where the lowest-order calculation ends and we can now discuss the corrections.\\

\begin{figure}[h]
\centering \mbox
{\hspace*{-1cm}\subfigure{\includegraphics[width=8.8cm]
{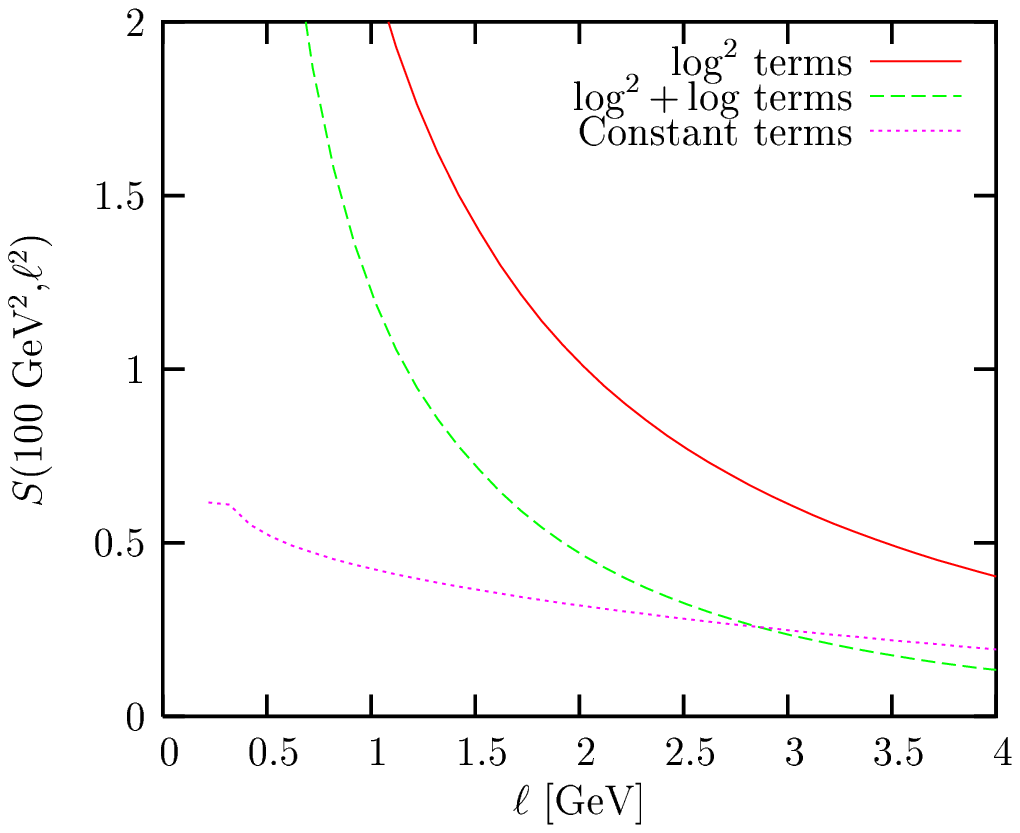}}\quad\quad
\hspace*{-1cm}\subfigure{\includegraphics[width=8.8cm]{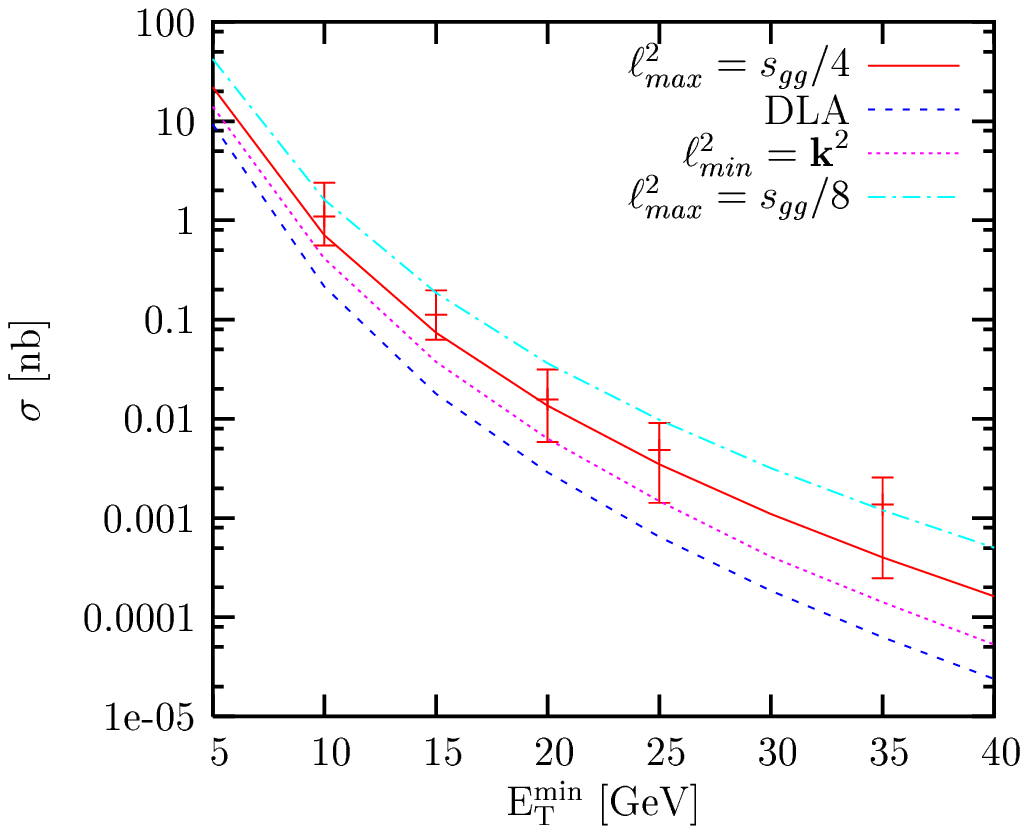}}}
\caption{\label{sudaterms} (a) Contribution of the double logs, double + single 
logs and constant terms to the Sudakov form factor as a function of the 
lower virtuality $\ell^2$ and for $E_T^{min}$=10~GeV. (b) Different 
prescriptions for the Sudakov form factor built up in the splitting function approximation: The full line is our reference curve with $\ell^2=s_{gg}$, $\mu^2=(\bk+\bk_i)^2$ and contains the double logs + single logs + constant terms contributions. \emph{DLA} includes only the double log contribution, $\ell^2$~ is the lower scale of the Sudakov form factor and $\mu^2$ is the upper scale.}
\end{figure}

The first one is the introduction of an impact factor that embeds the partons in the proton and that prevents the infrared divergence of the amplitude. It is a soft quantity that depends on the momenta of the incoming gluons and we have to fit it to data. In our study, we have used two kinds of impact factors. The first one is a very simple impact factor where the parton wave functions are parametrized on the light cone~\cite{Cudell:1993ui} and the second one is based on the unintegrated gluon density fitted to HERA data on $F_{2p}$~\cite{Ivanov:2000cm}. This second impact factor includes a dependence on the longitudinal momentum fraction $x$ lost by the proton. If we fix the Sudakov form factor, we can see on Fig.~\ref{impact} that the impact factor including the $x$-dependence has a slope in better agreement with the data points. The different choices give us a factor of 2 to 6 of uncertainty on the cross section. The largest uncertainty in central exclusive production comes from the Sudakov form factor which resums the large double logarithms coming from additional virtual-gluon loops. However, there are some problems with the exact structure of this factor. Firstly, the single-log contributions are not completely calculated in the exclusive case. Secondly, because exponentiation has been proven only for double logs and some single logs,  we do not know how to deal with the constant terms. These constant terms can be large and are not negligible as shown in Fig.~\ref{sudaterms}.a.
The main effect of the Sudakov form factor is to suppress strongly the cross section, but the factor ranges from 100 to 1000, depending on the prescription and on the scale involved, as shown in Fig.~\ref{sudaterms}.b. Furthermore, the Sudakov form factor changes the mean value of the momentum in the loop and we now have $<\bk>$ of the order of 1~GeV instead of 500~MeV. We do not find that the calculation becomes really perturbative as the shift in $<\bk>$ is not large. All these observations show that the Sudakov form factor is not under theoretical control and that it introduces a factor 10 of uncertainty on the cross section. 

The last piece is the gap survival probability. As was said previously, the daughter proton and antiproton can rescatter and the rescatterings can lead to large corrections, as we are close to the Black Disc limit. At the Tevatron energy, we used a gap survival probability equal to 10$\%$ but different calculations predict anything from few $\%$ to 35$\%$ of the cross section~\cite{Godbole:2008ny,Frankfurt:2006jp}. The different models introduce at least a factor of 3 of uncertainty in the calculation.
\section{Conclusion}
We have shown that the calculation of central exclusive dijet production is strongly sensitive to the corrections and the current uncertainties are large. If we combine all the previous numbers we find a factor 120 of uncertainty that leads to a large band of possible theoretical predictions on the dijet cross section and hence on the Higgs production cross section. This huge factor comes from the different prescriptions for the Sudakov form factor and the errors on the fit leading to the impact factor and may be reduced using the CDF data on exclusive dijet production. 
\section*{Acknowledgements} We thank V. Khoze for useful information
on implementing the transverse energy cut-off at the partonic level,
and acknowledge exchanges with M. Ryskin, E. Luna and K. Goulianos.

\end{document}